\documentstyle[aps,epsf]{revtex}  

\begin{document}        

\baselineskip 14pt
\title{
Probing CKM via 
Semileptonic B Decays
at the $\Upsilon$(4S)
}
\author{K. Kinoshita}
\address{University of Cincinnati}
\maketitle

\begin{abstract}
Semileptonic decays of $B$ mesons may be used in many different 
ways to probe the CKM matrix.
Presented here are several recent results from the CLEO~II experiment, all
involving the analysis or utilization of semileptonic
$B$ decays.

\end{abstract}

%
\section{Introduction}               
\begin{figure}[ht]	
\centerline{
\epsfxsize 2.5 truein \epsfbox{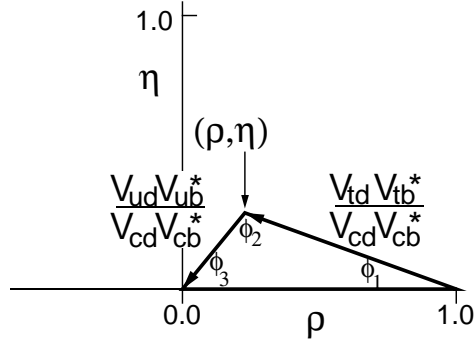}
}   
\caption[]{
\label{fig:uni}
\small ``Unitarity triangle'' in $\rho-\eta$ complex plane.}
\end{figure}

%
One of the major goals for the study of weak decays of the $B$ meson
is the measurement of
several third generation elements in the Cabibbo-Kobayashi-Maskawa 
(CKM) matrix
with sufficient accuracy to determine whether the matrix is unitary.
The requirement of unitarity constrains the number of free
parameters in this matrix to four.
The four are often shown in the explicit form 
\begin{center}
 ${\pmatrix{1-{\lambda^2\over 2} & \lambda & {\lambda^3A(\rho-i\eta )} \cr
                   -\lambda & 1-{\lambda^2\over 2} & {\lambda^2A} \cr
                   {\lambda^3A(1-\rho-i\eta)} & -\lambda^2A & 1} }$.
\end{center}
One condition of unitarity is that the scalar product of any column with
the complex conjugate of any other equals zero which, when
applied to the first and 
third columns, gives
\begin{eqnarray*}
0&=&V_{ub}^*V_{ud}+V_{cb}^*V_{cd}+V_{tb}^*V_{td}
\end{eqnarray*} 
When divided by the second term
\begin{eqnarray*}
0= {V_{ub}^*V_{ud}\over V_{cb}^*V_{cd}}+1+
{V_{tb}^*V_{td}\over V_{cb}^*V_{cd}}
\end{eqnarray*}
and presented explicitly in terms of the parameters $\rho$ and $\eta$,
this constraint may be illustrated graphically as a 
triangle in the complex plane, as shown in Figure~\ref{fig:uni}.
This triangle is known as a ``unitarity triangle.''

\begin{figure}[ht]	
\centerline{
\epsfxsize 2.3 truein \epsfbox{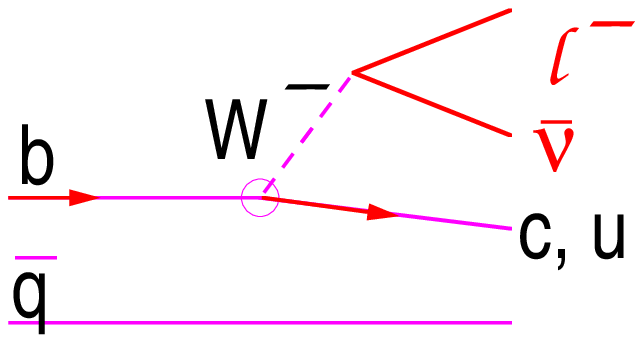}
\epsfxsize 3.0 truein \epsfbox{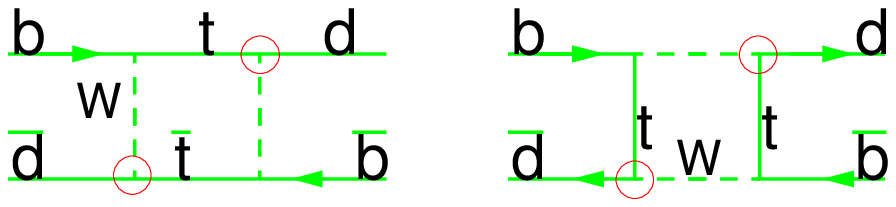}
}   
\vskip 5pt
\caption[]{
\label{fig:tree}
\small Processes used to measure CKM elements.
(left) Semileptonic decay: $b\rightarrow c$ probes
$|V_{cb}|$ and $b\rightarrow u$ is sensitive to $|V_{ub}|$.
(center, right) Examples of ``box'' diagrams describing $B^0$ mixing,
sensitive to $|V_{td}|$. }
\end{figure}

The least known of these CKM elements are those involving $b$ or $t$, and
by studying the partial rates to selected $B$ decay modes,
one is effectively measuring the lengths of the sides of this triangle.
In this talk I report recent results from the CLEO~II experiment for 
$\bar B\rightarrow D\ell^-\bar\nu$,
$\bar B\rightarrow (\pi/\rho/\omega) \ell^-\bar\nu$,
and
$B^0$ mixing, which are sensitive to the elements
$|V_{cb}|$, $|V_{ub}|$, and $|V_{td}|$ respectively, through the processes
illustrated in Figure~\ref{fig:tree}.

\section{Data}
%
%

The data were collected with the CLEO~II detector\cite{detector} at 
the Cornell Electron Storage Ring (CESR).
Most of the results reported here were obtained using a sample
of $e^+e^-$ annililation events collected during the
period 1990-5, with an integrated 
luminosity 3.1~fb$^{-1}$ on the ${\Upsilon (4S)}$
resonance ($\sqrt s=10.58$~GeV) and 1.13~fb$^{-1}$ at a center-of-mass 
energy which is
lower by 60~MeV (continuum).

The proximity of the $\Upsilon$(4S) resonance to the $B\bar B$ 
mass threshold and the
$B$ meson's lack of spin make the process $e^+e^-\rightarrow\Upsilon$(4S) 
particularly amenable to the experimental study of $B$ decays.
Because the resonance is only 20~MeV above the $B\bar B$ mass threshold, 
nearly all resonance events decay exclusively to $B\bar B$, with
roughly equal numbers to charged and neutral mesons.
At the same time, each $B$ is very slow, $\beta\gamma\approx 0.06$, 
so the assumption that the lab frame is the $B$ center of mass
is a useful approximation in many cases.
Because the $B$ is spinless, the angular distributions of decay
products from one $B$ are uncorrelated with those from the other.
As a consequence, the event as a whole tends to have a spherical rather than 
jetty shape, and the decay of each $B$ may be treated as an event with
energy $\sqrt s$/2 in its center of mass.

The low speed of the $B$ in the lab frame enables the use of a method 
known as ``partial reconstruction,'' where an exclusive decay mode is
identified through one or more
particles that are detected (Y) and one that is not (X).  
As all of the analyses reported in this talk exploit this method to some
degree, I present here a brief outline to which later discussions
will refer.
The requirements of energy-momentum conservation
\begin{eqnarray*}
E_X=E_B-E_Y,\ \ 
\vec{p}_X = \vec{p}_B-\vec{p}_Y
\end{eqnarray*}
which lead explicitly to the mass constraint
\begin{eqnarray}
M_X^2&=&E_X^2-{|\vec{p}_X|}^2 \nonumber\\ 
&=&(E_B-E_Y)^2-|\vec{p}_B|^2-|\vec{p}_Y|^2+2|\vec{p}_B||\vec{p}_Y|
{\rm cos}\theta_{B\cdot Y}
\label{eqn:mnu2}
\end{eqnarray}
form the basis for this method.
In the equation~(\ref{eqn:mnu2}) only the quantity ${\rm cos}\theta_{B\cdot Y}$
is unknown.
In one approach, the lab frame is taken to be the center 
of mass ($|\vec{p}_B|=0$) so that
the ``missing mass squared'' is approximated as
\begin{eqnarray}
M_X^2\approx (E_B-E_Y)^2-|\vec{p}_Y|^2.
\label{eqn:mnu2a}
\end{eqnarray}
Its distribution peaks broadly around the expected value.
In another approach, one examines the distribution of 
${\rm cos}\theta_{B\cdot Y}$: 
\begin{eqnarray}
{\rm cos}\theta_{B\cdot Y}={M_B^2+M_Y^2-M_X^2-2E_BE_Y\over 
2|\vec{p}_B||\vec{p}_Y|}
\label{eqn:cosby}
\end{eqnarray}
where one can expect correctly reconstructed candidates
to give physical values, $|{\rm cos}\theta_{B\cdot Y}|<1$.
\section{$|V_{cb}|$ via $\bar B\rightarrow D\ell^-\bar\nu$}

From a theoretical point of view, the relationship of {$|V_{cb}|$ to
the partial width $\Gamma(\bar B\rightarrow D\ell^-\bar\nu)$ 
is the product of the relevant couplings, kinematic
terms, and a hadronic form factor $F_D$, which is a function of
$w\equiv{m_B^2+m_D^2-q^2\over 2m_Bm_D}\ (1.00<w<1.59)$:
\begin{eqnarray*}
{d\Gamma\over dw}&=&{G_F^2|V_{cb}|^2\over 48\pi^3}(m_B+m_D)^2m_D^3(w^2-1)^{3/2}{F_D(w)^2}\\
\end{eqnarray*}
$F_D$ is usually expressed in a form with one or two free parameters, 
for example the quadratic form
\begin{eqnarray}
{F_D(w)\over F_D(1)}&\approx &1-\rho_D^2(w-1)+c_D(w-1)^2
\label{eqn:fd}
\end{eqnarray}

Experimentally, the measured estimator $\tilde w$ is only a fair
approximation to $w$ because the neutrino is not well measured,
so the shape of the original (root) distribution is distorted.
To regain the root distribution we measure the rate 
$\delta\tilde\Gamma_i(\tilde w)$ 
in ten bins and unfold via an
efficiency matrix $\epsilon_{ij}$: 
$\delta\Gamma_i=\epsilon_{ij}^{-1}\delta\tilde\Gamma_j$.
The unfolded distribution is then fitted for  $|V_{cb}|$ and the shape  
of $F_D$, using several parametrized forms of $F_D$.

At the $\Upsilon$(4S), the decay is reconstructed inclusively to
obtain the product branching fractions
\begin{eqnarray*}
{\cal B}(\Upsilon(4S)\rightarrow B^0\bar B^0){d{\cal B}\over d\tilde w}(
\bar B^0\rightarrow D^+\ell^-\bar\nu)&
\equiv &{f_{00}{dB_0\over d\tilde w}}\\
{\cal B}(\Upsilon(4S)\rightarrow B^+B^-){d{\cal B}\over d\tilde w}(
B^-\rightarrow D^0\ell^-\bar\nu)&
\equiv &{f_{+-}{dB_-\over d\tilde w}}
\end{eqnarray*}
where $f_{00}$($f_{+-}$) is the fraction of charged (neutral) events,
and $f_{00}+f_{+-}=1$.
These are unfolded to get root differential branching fractions,
which are then related to the differential width by
\begin{eqnarray*}
{d\Gamma\over dw}&=&{1\over \tau_-}{dB_-\over dw}={1\over \tau_0}
{dB_0\over dw}\\
&=& {1\over \tau_-}{\left(f_{+-}{dB_-\over dw}\right)}+
    {1\over \tau_0}{\left(f_{00}{dB_0\over dw}\right)}
\end{eqnarray*}
Measured in this way, $d\Gamma/dw$ is independent of $f_{00}$, 
which is not well measured.

Using 3.1~fb$^{-1}$ of CLEO data, candidates consist of an 
identified electron or muon and a candidate for $D^0\rightarrow K^-\pi^+$
or $D^+\rightarrow K^-\pi^+\pi^+$.
For each mode the distribution in the quantity $\cos\theta_{B\cdot D\ell}$ 
(described by equation~(\ref{eqn:cosby})) is examined.  
Backgrounds include candidates containing a fake $D$ or lepton, 
$D$ and $\ell$ from opposite $B$'s, continuum events,
and $D-\ell$ combinations from $B$ decay modes such as
$B\rightarrow DX\tau\nu(\tau\rightarrow\ell X)$ or
$B\rightarrow D_sD(D_s\rightarrow\ell X)$.
For each mode the distribution remaining after these background subtractions 
is shown in Figure~\ref{fig:Dlnu1} and includes,
in addition to the signal mode, large contributions from modes
$B\rightarrow DX\ell\nu$ where $DX$ originate with $D^*$, $D^{**}$, or
nonresonant $D^{(*)}\pi$ states.
To extract the distribution in $\tilde\omega$ of
$B\rightarrow D\ell\nu$, the  $\cos\theta_{B\cdot D\ell}$ distribution 
is fitted to a sum of simulated shapes for these
modes, in each of the ten bins of $\tilde\omega$
(Figure~\ref{fig:Dlnu2}).
This distribution is then unfolded to obtain the root distribution,
which is fitted to several theoretical forms\cite{boydcaprini}.

\begin{figure}[ht]	
\centerline{\epsfxsize 3.0 truein \epsfbox{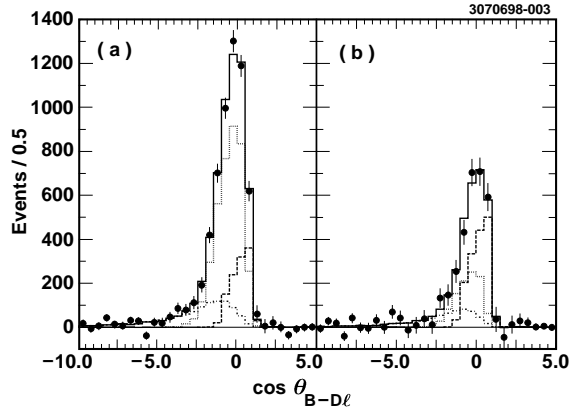}}   
\caption[]{
\label{fig:Dlnu1}
\small Distributions in cos$\theta_{B\cdot D\ell}$ for (a) 
$B\rightarrow D^0X\ell\nu$ and (b) $B\rightarrow D^+X\ell\nu$,
data (solid circles) with results from fitting to simulations (solid
histogram), which include
contributions from $B\rightarrow D\ell\nu$
(dashed), $B\rightarrow D^*\ell\nu$ (dotted), and
$B\rightarrow D^{**}\ell\nu+D^{(*)}\pi\ell\nu$ (dash-dotted).}
\end{figure}

\begin{figure}[ht]	
\centerline{\epsfxsize 2.5 truein \epsfbox{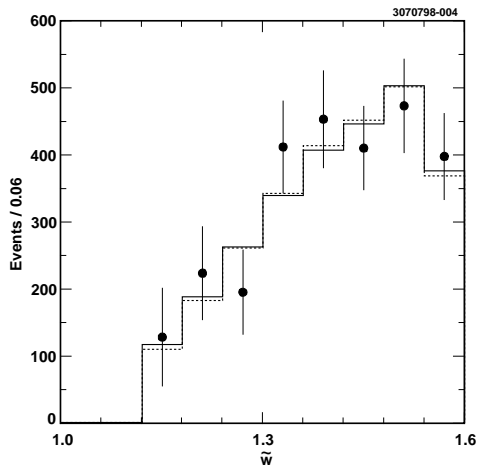}}   
\vskip 0 cm
\caption[]{
\label{fig:Dlnu2}
\small The sum of $B^-\rightarrow D^0\ell^-\bar\nu$ and 
$\bar B^0\rightarrow D^+\ell^-\bar\nu$ yields as a function of
$\tilde w$, for the data (solid circles) and from the best fit linear form
factor (dashed histogram) or dispersion relation inspired form factor
of Boyd {\it et al.} (solid histogram).}
\end{figure}

The results for partial width and branching fractions are
\begin{eqnarray*}
\Gamma(B\rightarrow D\ell\nu)&=&(14.1\pm 1.0\pm 1.2){\rm ns}^{-1}\\
{\cal B}(B^-\rightarrow D^0\ell^-\bar\nu)&=&(2.32\pm 0.17\pm 0.20)\%\\
{\cal B}(B^0\rightarrow D^+\ell^-\bar\nu)&=&(2.20\pm 0.16\pm 0.19)\%
\end{eqnarray*}
These are combined with a previous CLEO result\cite{Dlnuprev} to obtain
\begin{eqnarray*}
\Gamma(B\rightarrow D\ell\nu)&=&(13.4\pm 0.8\pm 1.2){\rm ns}^{-1}\\
{\cal B}(B^-\rightarrow D^0\ell^-\bar\nu)&=&(2.21\pm 0.13\pm 0.19)\%\\
{\cal B}(B^0\rightarrow D^+\ell^-\bar\nu)&=&(2.09\pm 0.13\pm 0.18)\%
\end{eqnarray*}
Using $F_D(1)=1.0$ gives
\begin{eqnarray*}
{|V_{cb}|=0.045\pm 0.006\pm 0.004\pm 0.005}
\end{eqnarray*}
where the errors are statistical, systematic, and theoretical, due to
uncertainty in $F_D(1)$.
This result is consistent with the most precise current value, 
$|V_{cb}|=0.0395\pm 0.0017$\cite{pdg}, obtained using the
decay $B\rightarrow D^*\ell\nu$.

\section{$|V_{ub}|$ via exclusive semileptonic decays}
At present, semileptonic $B$ decays are the only ones which are able to
provide measurements of $|V_{ub}|$.
Experimentally, the measurement is difficult because the rate is small and
the decays can be studied only in limited kinematic regions
where they are not overwhelmed by backgrounds from the much more abundant
$b\rightarrow c$ decays.
The difficulty is compounded by the
spread among theoretical models
which provide the relationship between measured rates and the matrix
element.
In this context, exclusive and inclusive semileptonic decays yield
measurements of $|V_{ub}|$ in somewhat complementary ways.
Presented here is a measurement using five exclusive modes and a method
that differs from that used in the previously published CLEO~II
result based on the same modes\cite{lkg}.

To identify an exclusive decay, a track is identified as an
electron or muon and combined with a $u$-hadron candidate 
($\rho^\pm\rightarrow\pi^\pm\pi^0,\ \rho^0\rightarrow\pi^+\pi^-,
\ \omega\rightarrow\pi^+\pi^-\pi^0,\ \pi^0\rightarrow\gamma\gamma,\ {\rm or}
\ \pi^\pm$).
In addition, a neutrino candidate is formed based on 
missing momentum and energy in the event.
The three-particle combination must be kinematically 
consistent with originating from a $B$ decay.
The principal background in the most significant kinematic regions
arises from continuum events, so requirements are designed to suppress
these strongly.
The following constraints, which are due to light quark symmetry, are applied:
\begin{eqnarray*}
\Gamma(\bar B^0\rightarrow\rho^+\ell^-\bar\nu)&=&
2\Gamma(B^-\rightarrow\rho^0\ell^-\bar\nu)\\
&=&2\Gamma(B^-\rightarrow\omega^0\ell^-\bar\nu)\\
\Gamma(\bar B^0\rightarrow\pi^+\ell^-\bar\nu)&=&
2\Gamma(B^-\rightarrow\pi^0\ell^-\bar\nu)
\end{eqnarray*}
The candidates are sorted by lepton momentum into three samples, [2.3-2.7],
[2.0-2.3], and [1.7-2.0] GeV/c.
A maximum likelihood fit is then performed for all three samples on 
distributions of the five modes in the quantity 
$\Delta E$, defined as the
candidate energy minus the beam energy and, where applicable, the 
invariant mass of the $\rho$ or $\omega$ candidate.
The fit accounts for contributions not only from the signal modes but also
from other decays of the type $b\rightarrow u\ell\nu$, $b\rightarrow c$
semileptonic decays, fake leptons, and continuum.
To allow for uncertainties in the contribution from $b\rightarrow c$
modes, each bin of lepton momentum is normalized separately,
although the normalization among distributions within each bin is common.
A total of twelve free parameters remain in the fit.
A projection onto the $\pi\pi$ invariant mass of the fit for 
the sum of $\rho$ modes in the highest 
lepton momentum bin is shown in Figure~\ref{fig:rholnu1}.
We obtain a result for the branching fraction 
${\cal B}(\bar B^0\rightarrow\rho^+\ell^-\bar\nu)$ assuming that
neutral and charged $B$'s are produced in equal abundance 
at the $\Upsilon$(4S).
From the branching fraction and the measured $B^0$ lifetime\cite{pdg}
we obtain the partial width and $|V_{ub}|$.
\begin{figure}[ht]	
\centerline{\epsfxsize 2.5 truein \epsfbox{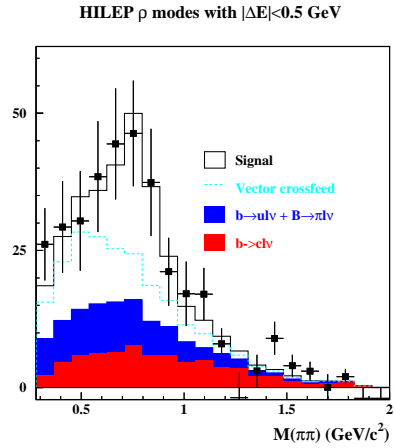}}   
\vskip -.2 cm
\caption[]{
\label{fig:rholnu1}
\small Projection of maximum likelihood fit onto $\pi\pi$ invariant
mass in the highest lepton momentum bin for the sum of $\rho$ modes.
For this plot a requirement of $|\Delta E|<500$~MeV has been made.}
\end{figure}

Both the reconstruction efficiency and extraction of $|V_{ub}|$ from
the measurement are somewhat model-dependent.
To estimate the degree of uncertainty due to models, the
analysis is repeated for each of five current models\cite{models}.
The resulting values are presented in Figure~\ref{fig:rholnu2}. 
For each result the average is given as the central value
and the model dependence error is taken to be half the full
spread among the five results.  
This measurement of $|V_{ub}|$ is comparable to that obtained via inclusive
semileptonic decays, $|V_{ub}|=(3.2\pm 0.8)\times 10^{-3}$\cite{pdg}.
These results are largely independent of our previously published 
result\cite{lkg}.
An average which takes into account the statistical and systematic 
correlations will be presented in the near future.

It may be possible to address the uncertainties due to model 
dependence through the measurement
of the decay rate as a function of $q^2$.
To investigate this possibility, the data were subdivided into 
three bins, [0-7], [7-14], and 
[14-21] (GeV/c$^2$)$^2$, the maximum allowed by current statistics.
The results for the partial widths $\Delta\Gamma(B\rightarrow\rho\ell\nu)$ are
\begin{eqnarray*}
(7.6\pm 3.1^{+0.9}_{-1.2}\pm 3.1)\times 10^{-5}{\rm ps}^{-1} &\ \ \ \ & 
0<q^2<7\ {\rm GeV}^2/c^4\\
(4.8\pm 2.9^{+0.7}_{-0.8}\pm 0.7)\times 10^{-5}{\rm ps}^{-1}&&
7<q^2<14\ {\rm GeV}^2/c^4\\
(7.1\pm 2.1^{+0.9}_{-1.1}\pm 0.6)\times 10^{-5}{\rm ps}^{-1}&&
14<q^2<21\ {\rm GeV}^2/c^4
\end{eqnarray*}
where the errors are statistical, systematic, and model spread,
respectively.
The highest of these bins has the smallest model variation.

\begin{figure}[ht]	
\centerline{
\epsfxsize 2.8 truein \epsfbox{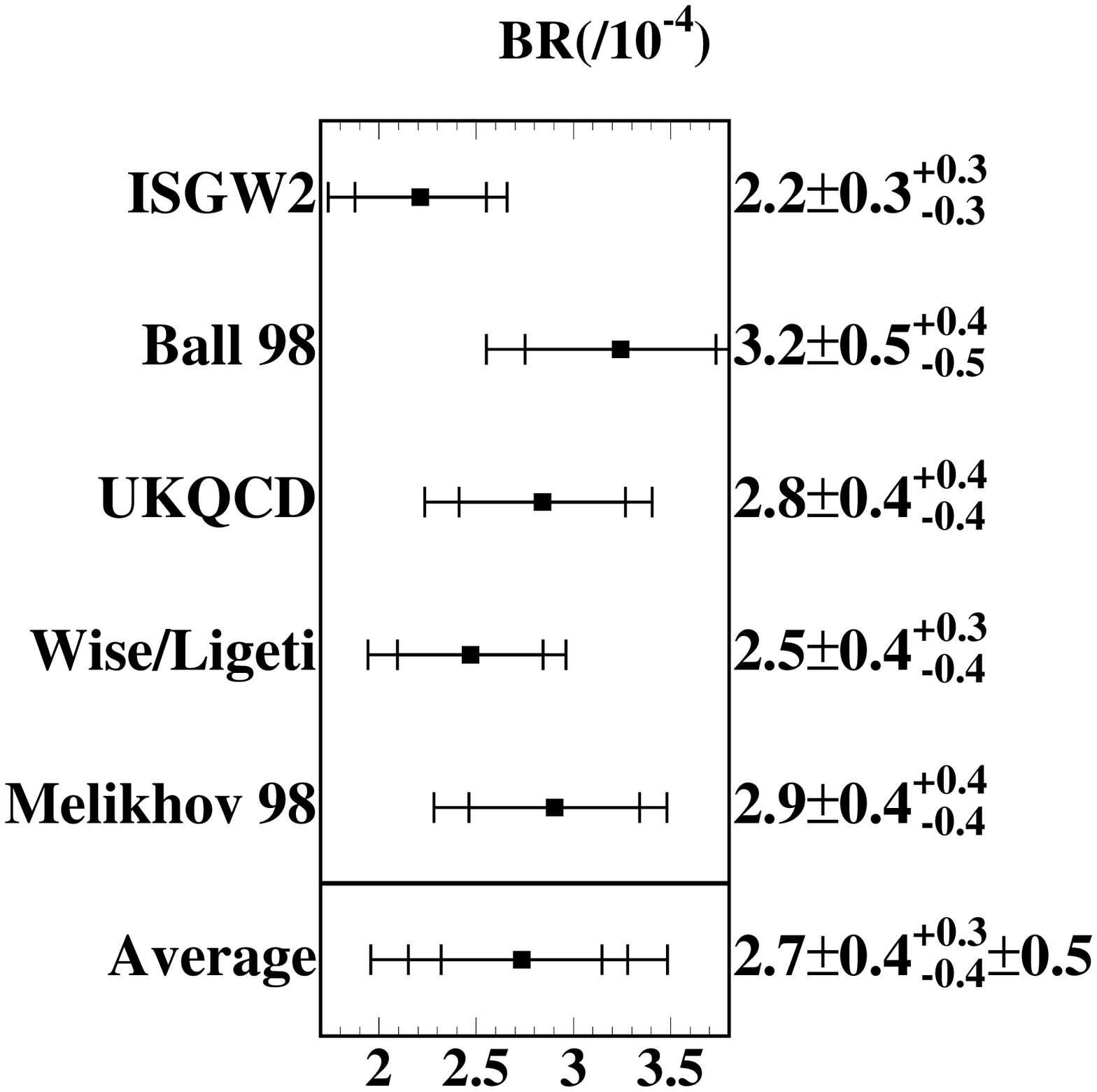}
\epsfxsize 2.8 truein \epsfbox{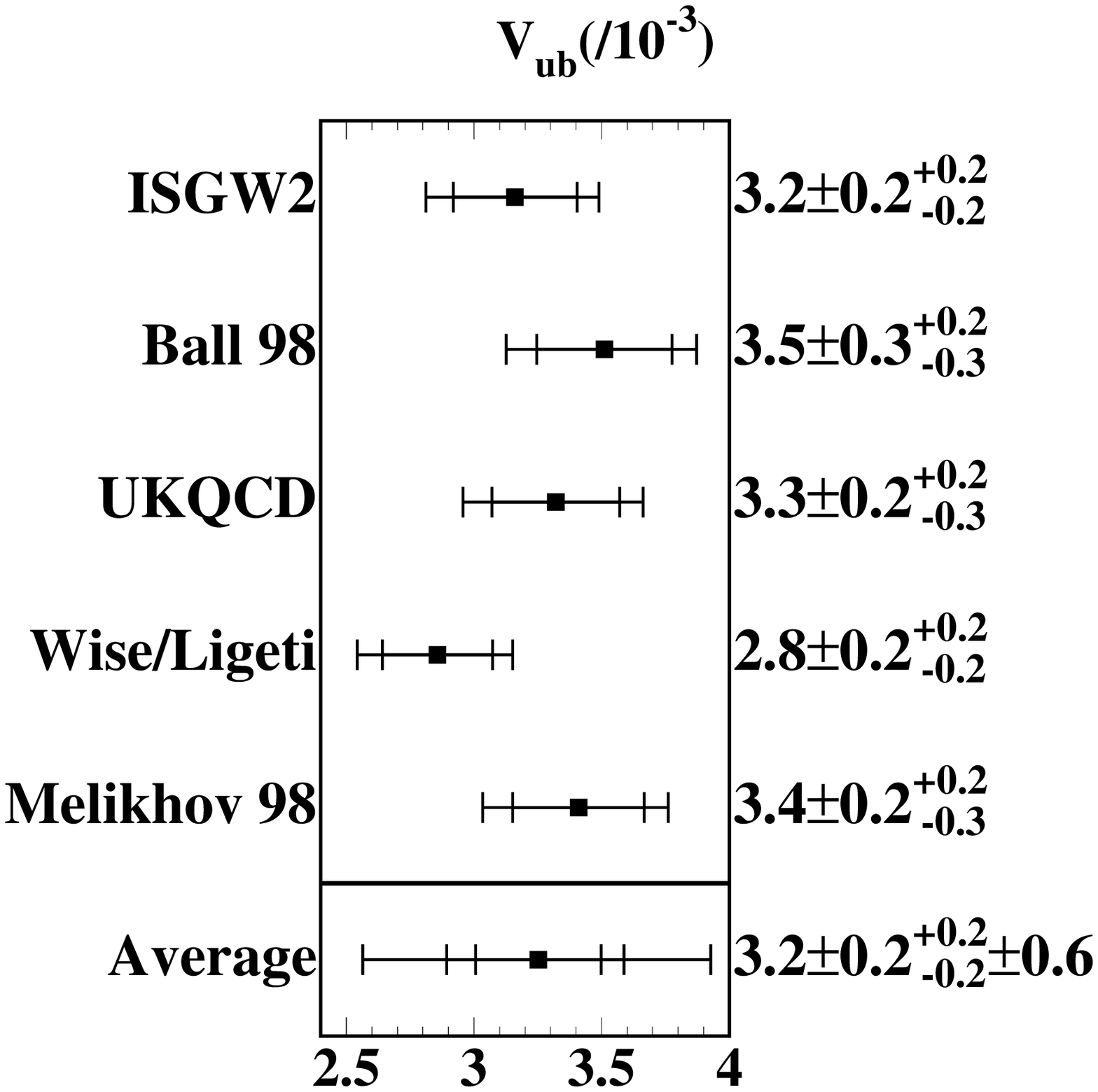}
}   
\vskip -.2 cm
\caption[]{
\label{fig:rholnu2}
\small Branching fraction ${\cal B}(\bar B^0\rightarrow\rho^+\ell^-\bar\nu)$ 
(left) and  $|V_{ub}|$ (right) for five models, with averages (preliminary). }
\end{figure}

\section{$|V_{\mit td}|$ via $B^0$ mixing}

Mixing occurs for neutral $B$ mesons through 
second order weak processes such as those represented in the box diagrams in
Figure~\ref{fig:tree}.
It causes the exponential decay
pattern of a population of $B$'s to contain an oscillatory component,
as is illustrated for a hypothetical case in Figure~\ref{fig:mix1}.
The oscillation rate $\omega$ is proportional to $|V_{td}|^2$,
and the fraction $\chi_d$ of an initial population of $B^0$ that eventually
decays as $\bar B^0$ may be expressed in terms of the product 
$\omega\tau\equiv x_d$
where $\tau$ is the lifetime:
\begin{eqnarray*}
1-\chi_d={2+x_d^2\over 2(1+x_d^2)}
\end{eqnarray*}

\begin{figure}[ht]	
\centerline{\epsfxsize 2.4 truein \epsfbox{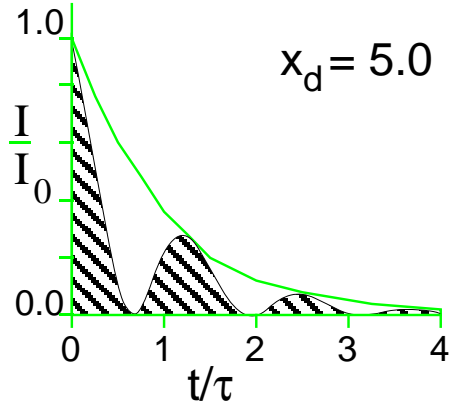}}   
\vskip -.2 cm
\caption[]{
\label{fig:mix1}
\small Illustration of mixing: $B$ decays observed given an initial
population of pure $B$, for $x_d=5.0$
}
\end{figure}

At the $\Upsilon$(4S), mixing is manifested by the presence of ``mixed''
events, where two $B$ decays of the same flavor, $B^0B^0$ or 
$\bar B^0\bar B^0$, are found in the same event.
In this case, $\chi_d$ is equal to the fraction of mixed
events among neutral $B$ events.

%
Presented here are two methods of tagging that enable the clean flavor
identification of both $B$'s in an event.
In tagging, the objective is to reconstruct one of the $B$'s sufficiently
to identify its flavor and to identify it as a neutral $B$.
The flavor of the second $B$ is then identified through a lepton
with high momentum, which originates predominantly from decays of
the type $B\rightarrow c\,\ell\nu$.
The tagging is accomplished through partial reconstruction of two
different decay modes:
\begin{enumerate}
\item {$\bar B^0\rightarrow D^{*+}{\ell^-}\bar\nu\ 
\{D^{*+}\rightarrow D^0{ \pi_s^+}\}$}, where two of the particles,
$\bar\nu,\ D^0$ are not detected but can be constrained due to the
kinematics of $B$ and $D^*$ decays:\\
\begin{eqnarray*}
E_{D^*}\approx E_{\pi_s}{M_{D^*}\over E_{\pi}^{CM}}
\equiv \tilde E_{D^*}\ \ \tilde{p}_{D^*}
\equiv \hat{p}_{\pi_s}\sqrt{\tilde{E}_{D^*}^2-m_{D^*}^2}\\
\Rightarrow \widetilde{M}_\nu^2=(E_B-E_\ell-\tilde{E}_{D^*})^2-
|\vec{p}_B-\vec{p}_\ell-\tilde{p}_{D^*}|^2
\end{eqnarray*}
\item {$\bar B^0\rightarrow D^{*+}{ (\pi^-/\rho^-)}\ 
\{D^{*+}\rightarrow D^0{ \pi_s^+}\}$}, where the $D^0$ is not
detected but can be fully constrained (except for a twofold
ambiguity) through energy and momentum conservation.
\end{enumerate}

To identify $\bar B^0\rightarrow D^{*+}{\ell^-}\bar\nu$, 
we select a high momentum ($>1.4$~GeV/c) electron or muon and
a soft track ($<0.190$~GeV/c) with the opposite charge.
For signal candidates, where the pair originate from the signal
decay as specified above, the distribution in $\widetilde{M}_\nu^2$
form a peak centered near  $\widetilde{M}_\nu^2=0$, as shown in
Figure~\ref{fig:mix2}.
The backgrounds are formed from random combinations of leptons and soft
tracks and may originate in continuum or $\Upsilon$(4S) events.
Continuum backgrounds may be estimated using our nonresonant data
sample, and backgrounds from  $B\bar B$ events are estimated via
Monte Carlo simulation.

\begin{figure}[ht]	
\centerline{\epsfxsize 2.5 truein \epsfbox{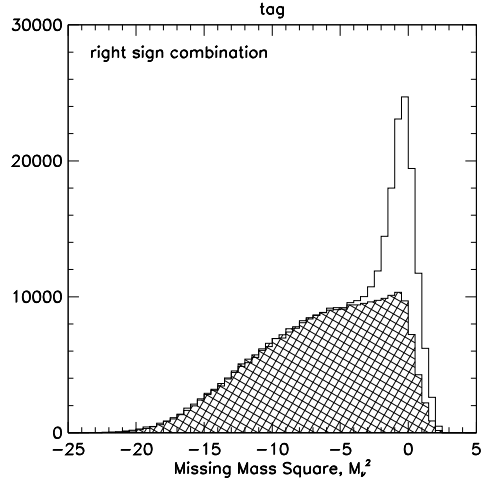}}   
\vskip -.0 cm
\caption[]{
\label{fig:mix2}
\small Distributions in $\widetilde{M}_\nu^2=0$ of candidates for
$\bar B^0\rightarrow D^{*+}{\ell^-}\bar\nu$,
data with continuum contribution subtracted (solid) and simulated
background (cross-hatched).
}
\end{figure}

Candidates are further sorted according to the momentum of any additional
leptons that are identified in the event and whether the additional
lepton has the opposite (``unmixed'') or same (``mixed'')
charge as the lepton of the tag.
For each sorted set, the number of tags is extracted according to
the same procedure.
The additional leptons are mainly from primary decays of the type
$b\rightarrow c(u)\ell^-\bar\nu$ and secondary decays $b\rightarrow
cX \{c\rightarrow s\ell^+\nu\}$.
Leptons from other sources, including hadrons misidentified as
leptons, secondary leptons from decays of $B$ to $\psi$, $\Lambda_c$,
$\tau$, and $\bar c$, and electrons from $\gamma\rightarrow e^+e^-$
and $\pi^0\rightarrow\gamma e^+e^-$ are accounted for and subtracted.
The net number in each bin is corrected for the detection efficiency
of the additional lepton.

The resulting lepton  momentum distributions are shown in
Figure~\ref{fig:mix3}.
Each is fitted to a sum of primary and secondary spectrum
shapes to extract the net number of primary leptons with mixed or
unmixed events.
The mixing parameter $\chi_d$ is straightforwardly derived from the
ratio of the two fitted numbers.

To reconstruct $\bar B^0\rightarrow D^{*+}(\pi^-/\rho^-),\ \{D^{*+}\rightarrow D^0{ \pi_s^+}\}$, we select
candidates consisting of a hard track or $\rho^\pm$ candidate 
($p_h>1.4$~GeV/c) and a soft
track with opposite charge.
These are required to be kinematically consistent with being the hard hadron
and the soft pion from $D^*$
originating with the
signal decay, and this constraint is sufficient to yield a high
signal relative to background.
Events containing tag candidates are then searched for a hard lepton
($p_l>1.4$~GeV/c) and sorted as an ``unmixed'' event if the hard hadron
and lepton have opposite charge and as a ``mixed'' event if they
have the same charge.
The mixing parameter $\chi_d$ is extracted by accounting for the various
correlated contributions to the mixed and unmixed samples, including
true mixing, continuum, and background candidates from $B^0$ and from $B^-$.

\begin{figure}[ht]	
\centerline{
\epsfxsize 2.8 truein \epsfbox{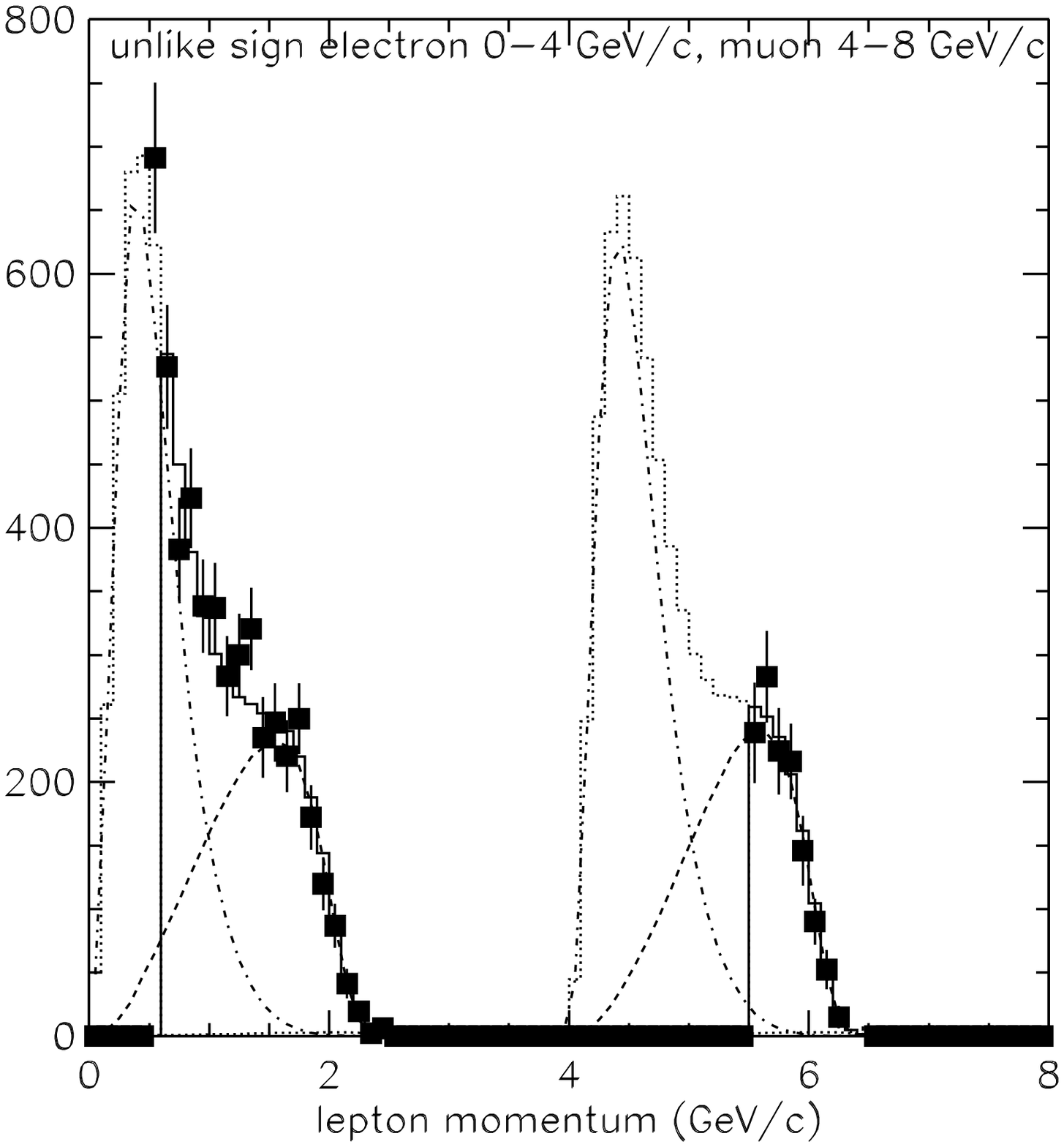}
\epsfxsize 2.8 truein \epsfbox{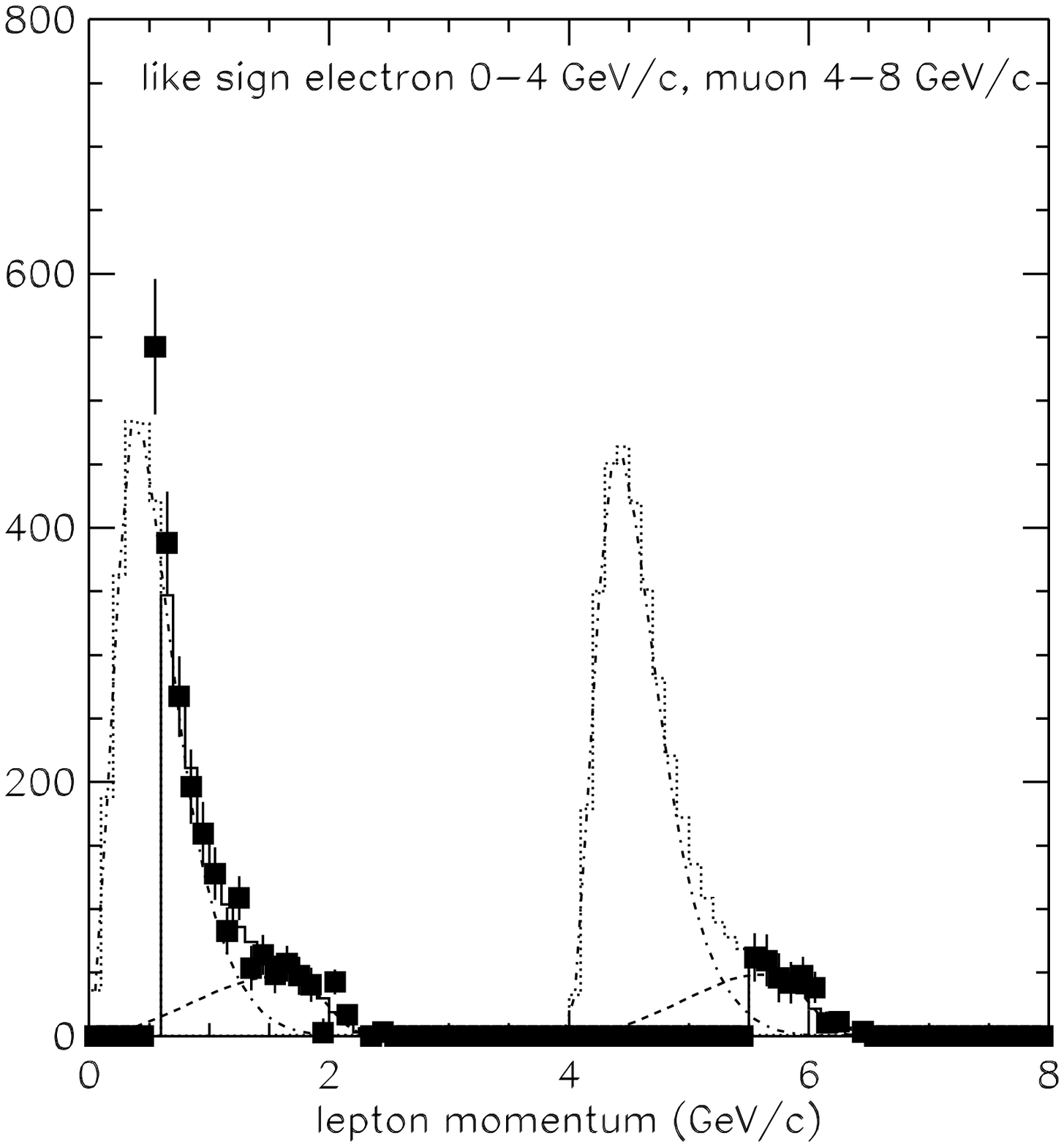}
}   
\vskip 8pt
\caption[]{
\label{fig:mix3}
\small Distributions of additional leptons in events containing 
$\bar B\rightarrow D^{*+}X\ell\bar\nu$, shown with fits to 
primary (dashed) and secondary (dash-dotted) spectra. 
Electron momentum is plotted in the region 0.0-4.0~GeV/c and \{muon
momentum+4.0~GeV\} is plotted in the region 4.0-8.0~GeV/c.
In the left plot the two leptons have opposite charge 
(unmixed),
and in the right plot they have the same charge (mixed).
}
\end{figure}

Preliminary results for both of these mixing analyses are presented here.
Using the tag of $\bar B\rightarrow D^{*+}\ell\bar\nu$ with
3.1~fb$^{-1}$ of data we find
\begin{eqnarray*}
\chi_d&=&0.187\pm 0.019\pm 0.007\\
x_d&=& 0.773\pm 0.048\pm 0.018\\
|V_{td}|&=& 0.0086\pm 0.0003\pm 0.0001\pm 0.0018
\end{eqnarray*}
where the third error on $|V_{td}|$ is due to uncertainties in the
theory.
The result from the $\bar B^0\rightarrow D^{*+}(\pi^-/\rho^-)$ tag
with 3.99~fb$^{-1}$ of data is
\begin{eqnarray*}
\chi_d=0.191\pm 0.026\pm 0.014
\end{eqnarray*}
Both analyses are comparable in statistical significance to the
best previous results but have significantly reduced systematic errors
and should result in improvements with larger data sets.
However, 
the determination of $|V_{td}|$ is currently limited by
theoretical uncertainties, which will need to be reduced before
these result in better values.

\section{Summary}
Semileptonic $B$ decays are important for the study of the third
generation CKM matrix elements and for investigating 
the unitarity of the matrix.
Recent measurements of the rates for $B\rightarrow D\ell\nu$,
$B\rightarrow (\rho/\omega/\pi)\ell\nu$, and $B^0$
mixing by the CLEO~II experiment have been presented here.
These rates are sensitive to $|V_{cb}|$, $|V_{ub}|$, and $|V_{td}|$,
respectively.
These measurements highlight the progress being made in the development
of experimental techniques and in the theoretical understanding of
the dynamics of heavy quark decay.

%

\end{document}